\documentclass[aps,prl,superscriptaddress, twocolumn, footinbib,longbibliography]{revtex4-1}
\usepackage{amsmath,amssymb}
\usepackage{graphicx}
\usepackage{dcolumn}
\newcolumntype{d}[1]{D{.}{.}{#1}}

\usepackage{bm}
\usepackage{epstopdf}
\usepackage{graphicx}
\usepackage{stmaryrd}
\usepackage{tikz}
\usepackage{lipsum}
\usepackage{pgf}
\usepackage{float}
\usepackage[colorlinks, linkcolor= blue, citecolor = blue, urlcolor=blue]{hyperref}
\usepackage{tabularx}
\usepackage[none]{hyphenat}

\usepackage{subfigure}

\definecolor{lightblue}{RGB}{0,170,255}

\newcommand{\appropto}{\mathrel{\vcenter{
			\offinterlineskip\halign{\hfil$##$\cr
				\propto\cr\noalign{\kern2pt}\sim\cr\noalign{\kern-2pt}}}}}

\usepackage[normalem]{ulem}

\newcommand{\om}{\iffalse}
\newcommand{\pd}[2]{\frac{\partial #1}{\partial #2}}

\newcommand{\ba}{\arraycolsep 0.3ex \begin{array}{rl}}
	\newcommand{\ea}{\end{array}}
\newcommand{\bc}{\begin{cases}}
	\newcommand{\ec}{\end{cases}}
\usepackage{color}
\newcolumntype{C}[1]{>{\centering\arraybackslash}p{#1}}

\begin{document}
\title{Unidirectional magneto-transport of linearly dispersing topological edge states}
\author{Zhanning Wang}
\affiliation{School of Physics, The University of New South Wales, Sydney 2052, Australia}	
\affiliation{ARC Centre of Excellence in Future Low-Energy Electronics Technologies, The University of New South Wales, Sydney 2052, Australia}	

\author{Pankaj Bhalla}
\affiliation{ARC Centre of Excellence in Future Low-Energy Electronics Technologies, The University of New South Wales, Sydney 2052, Australia}
\affiliation{Beijing Computational Science Research Center, Beijing 100193, China}

\author{Mark Edmonds}
\affiliation{ARC Centre of Excellence in Future Low-Energy Electronics Technologies, Monash University, Clayton, Victoria 3800}

\author{Michael S. Fuhrer}
\affiliation{ARC Centre of Excellence in Future Low-Energy Electronics Technologies, Monash University, Clayton, Victoria 3800}

\author{Dimitrie Culcer}
\affiliation{School of Physics, The University of New South Wales, Sydney 2052, Australia}
\affiliation{ARC Centre of Excellence in Future Low-Energy Electronics Technologies, The University of New South Wales, Sydney 2052, Australia}

\date{\today}
\begin{abstract}
Quantum spin-Hall edges are envisaged as next-generation transistors, yet they exhibit dissipationless transport only over short distances. Here we show that in a diffusive sample, where charge puddles with odd spin cause back-scattering, a magnetic field drastically increases the mean free path and drives the system into the ballistic regime with a Landauer-Buttiker conductance. A strong non-linear non-reciprocal current emerges in the diffusive regime with opposite signs on each edge, and vanishes in the ballistic limit. We discuss its detection in state-of-the-art experiments.
\end{abstract}
\maketitle

Quantum spin-Hall insulators are a novel class of materials hosting gapless, topologically protected, counter-propagating edge states. These have opposite spin polarizations and exhibit strong spin-momentum locking due to the dominant role of the spin-orbit interaction \cite{Ref_9,Ref_54,Ref_31,Ref_38,Ref_45,Ref_46,Ref_35,Ref_7,Ref_15}. Time-reversal symmetry ensures edge states come in Kramers doublets, which cannot be back-scattered by time-reversal invariant perturbations \cite{Ref_59,Ref_60,Ref_20,Ref_23,Ref_24,Ref_61,Ref_65,Ref_66}. Materials possessing topological edge states include topological insulators such as $\mathrm{HgTe}$ and $\mathrm{Bi}_2\mathrm{Se}_3$, Weyl semimetals such as $\mathrm{WTe}_2$, and Dirac semimetals such as $\mathrm{Na}_3\mathrm{Bi}$ \cite{Ref_36,Ref_49,Ref_3,Ref_8,Ref_31,Ref_5,Ref_58,Ref_50}.

A ballistic edge has a longitudinal conductance of $e^2/h$ at low temperature, a fact that has led to proposals for using topological edge states as building blocks for next-generation transistors, exploiting electrically tunable topological phase transitions \cite{Ref_72}. Nevertheless, following the experimental discovery of topological edge states, it has emerged that puddles with odd numbers of charges, which exist inherently in the host materials due to doping disorder in fabrication, can act as effective magnetic impurities that back-scatter the edge states and significantly reduce their mobility \cite{Ref_10,Ref_20,Ref_14,Ref_23,Ref_63,Ref_37}. This may explain why ballistic conductance has only been observed over spatial scales of the order of 50 $\mathrm{nm}$ \cite{Ref_3,Ref_8,Ref_10,Ref_49,Ref_56,Ref_58,Ref_52,Ref_19}. Whereas initial studies focused on the Kondo effect, the Kondo temperature in current samples is expected to be negligibly small \cite{Ref_43, Ref_30, Ref_28}, while other aspects of transport remain poorly understood \cite{Ref_31,Ref_16,Ref_18,Ref_22,Ref_21,Ref_53,Ref_54,Ref_55,Ref_66,Ref_67}. The unexpectedly large resistance of topological edge states has emerged as a fundamental question and an obstacle in the development of topological transistors \cite{Ref_11, Ref_13}. Bearing in mind the role of magnetic impurities, the first step in overcoming this problem is understanding edge magneto-transport in the presence of puddles. This includes the identification of non-reciprocal currents, since non-linear response probes interactions that are difficult to access in linear response, due to constraints imposed by mirror symmetry and Onsager relations \cite{Ref_33,Ref_34,Ref_42,Ref_48,Ref_49}. The rich physics underlying non-linear phenomena \cite{Ref_6, Ref_5, Ref_68, Ref_69, Ref_70, Ref_71} has been manifest in recent discoveries such as Hall effects in time-reversal invariant systems, as well as in unexpected features of topological edges, such as a large uni-directional magneto-resistance at zero magnetic field \cite{Ref_3, Ref_5, Ref_48, Ref_49}.

\begin{figure}[t!]
\includegraphics[width=\columnwidth]{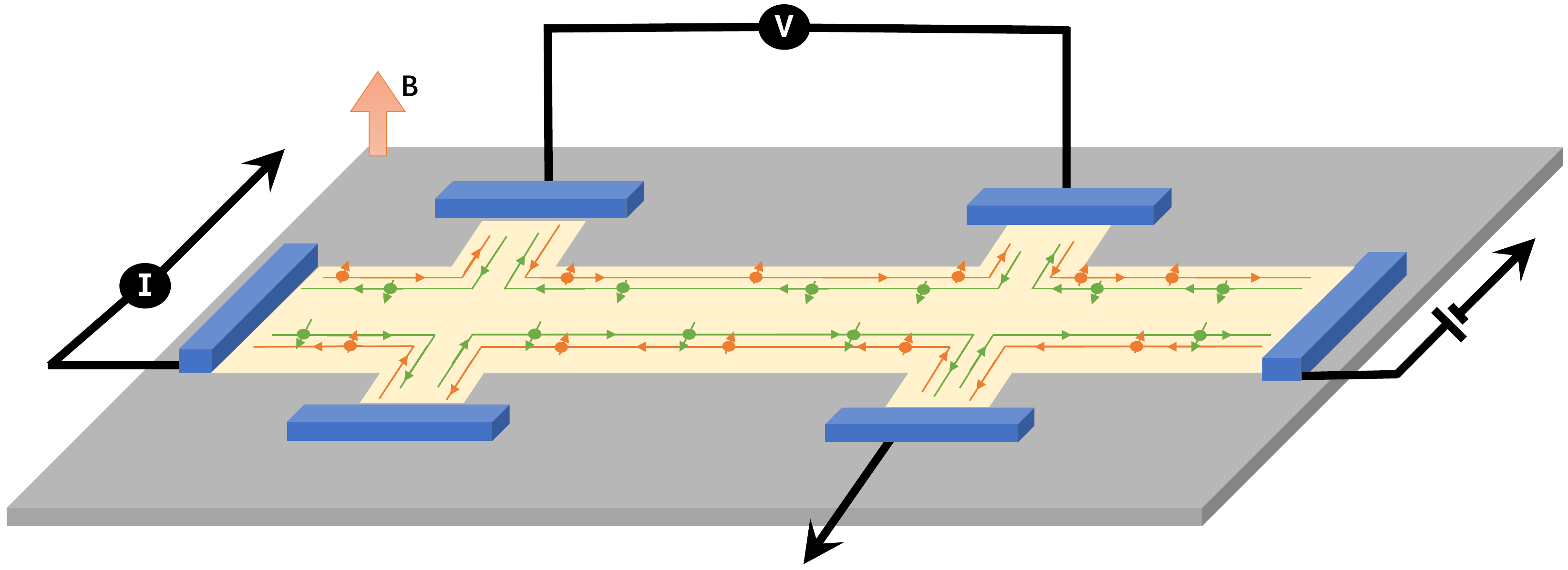}
\caption{Experimental setup in a perpendicular magnetic field $\bf{B}$. The current is measured across the sample, while the voltage can be measured at two different terminals along one side. On the opposite side one terminal is grounded. Spin-up electrons are shown in orange, spin-down electrons in green.}
\label{Figure: Experimental set-up}
\end{figure}

In this article we demonstrate that a magnetic field has a drastic effect on both the linear and non-linear response of topological edge states: (i) It enhances the mean free path $l$ by orders of magnitude without opening a gap, eventually driving the system into the ballistic regime; (ii) By breaking mirror symmetry the magnetic field enables a strong unidirectional non-linear electrical response in the diffusive regime. The direction of the current is determined by the magnetic field and the spin quantization axis, and it has a different sign on each edge. Interestingly, the non-reciprocal current vanishes in the ballistic regime. This reflects the fact that, once magnetic impurity scattering is surmounted, the only remaining magnetic interaction is the Zeeman interaction with the out of plane field, which can be gauged away. Whereas a complete description of charge puddles is beyond the scope of this work, modelling the puddles as magnetic impurities is a simple way of capturing the physics that governs their magnetoresistance, which is in excellent agreement with experiment \cite{Ref_52}.

\twocolumngrid
\begin{figure*}[!t]
\centering
\includegraphics[width=2\columnwidth]{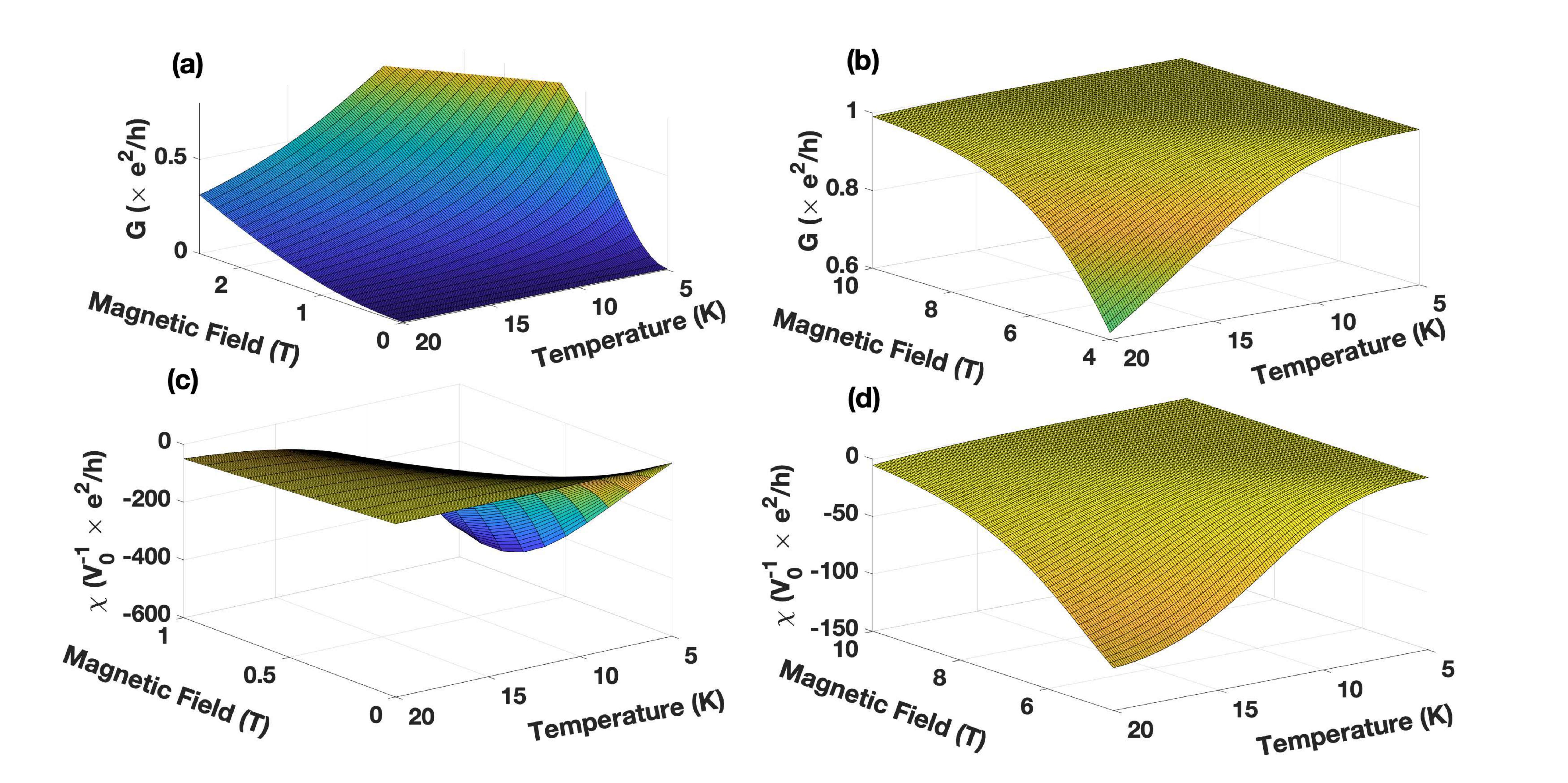}
\caption{Linear and non-linear response for a sample with $d=1000\mathrm{nm}$; $v_0 = 10^3 \mathrm{V}$ is a scaling factor. (a) The conductance in the diffusive regime. When the magnetic field is small, the conductance increases as back-scattering is suppressed and the mean free path increases. (b) The conductance in the ballistic limit, when the mean free path is larger than the size of the sample, the conductance will not change, converging to to $e^2/h$. (c) The nonlinear response function $\chi$ in the diffusive regime. When the magnetic field is small, the mean free path is much shorter than the size of the sample, and the non-linear response increases as a function of $\mathrm{B}$. (d) The nonlinear response function $\chi$ in the ballistic regime. At large magnetic fields the mean free path exceeds the size of the sample, causing $\chi$ to decrease in the ballistic regime and eventually vanish.}
\label{Figure: Four-panel conductivity plots}
\end{figure*}

Referring to the set-up shown in Fig.~\ref{Figure: Experimental set-up}, our main results are summarised in Fig.~\ref{Figure: Four-panel conductivity plots}. The current in the channel will be denoted by $I$ and the potentials of the left and right electrodes by $V_L$, $V_R$ respectively. We define the conductance $G$ and the non-linear electrical response function $\chi$ by $I = G(V_L - V_R) + \chi (V_L - V_R)^2$. In Fig.~\ref{Figure: Four-panel conductivity plots}-(a) and Fig.~\ref{Figure: Four-panel conductivity plots}-(b) we have plotted the conductance $G$ as a function of the applied out-of-plane magnetic field $B$ at small and large values of $B$, where small and large are quantified below. It is seen that $G$ increases with $B$ and eventually reaches the quantized Landauer-Buttiker value of $e^2/h$, indicating that the system reaches the ballistic limit. This opens up the exciting possibility of using a ferromagnet with an out-of-plane magnetization as a practical method to increase the mean free path and to study transport in the ballistic regime. The ferromagnet could couple to the impurities either via a magnetic field or through the exchange interaction. Next, Fig.~\ref{Figure: Four-panel conductivity plots}-(c) and Fig.~\ref{Figure: Four-panel conductivity plots}-(d) show the non-linear electrical response function $\chi$ at small and large magnetic fields respectively. At small $B$, $\chi$ increases with $B$, but in contrast to the Ohmic term the non-linear signal reaches a maximum beyond which it decreases, tending to zero as the system reaches the ballistic regime. This vanishing response is a characteristic of the Dirac cone, indicating that the non-linear response is a probe of the edge state dispersion, and is a unique experimental signature reflecting chiral conduction in the TI. To generate and detect the second-order response at low-frequency it is sufficient to use an oscillator with angular frequency $\omega$ and read off the signal at 2$\omega$.

We focus on $\mathrm{Na}_3\mathrm{Bi}$ as a prototype material, motivated by the observations that ultra-thin films of $\mathrm{Na}_3\mathrm{Bi}$ have a band-gap of $\geq$300 meV \cite{Ref_25}, much greater than $k_BT$ at room temperature, are robust to layer-number fluctuations caused by imperfect growth \cite{Ref_57}, exhibit an electrically driven topological phase transition\cite{Ref_4}, and show clear evidence of edge transport over millimetre distances, as well as a giant negative magneto-resistance \cite{Ref_52}. Our model also applies to topological insulators with inversion symmetry such as Bi$_2$Se$_3$. Materials without inversion symmetry, such as WTe$_2$, exhibit a positive magneto-resistance and a position-dependent spin quantization axis, so they fall outside our scope. 

Considering a sample of finite size $d$ a magnetic field ${\bm B} \parallel \hat{\bm z}$ is applied out of the plane. The full Hamiltonian $H$ can be written as $ H = H_0 + H_Z + V + U + U_Z$, where the band Hamiltonian $H_0 = \hbar v_F k_x \sigma_z $ represents the edge state dispersion of Na$_3$Bi; $H_Z = g_0 \mu_B B \sigma_z$ is the Zeeman interaction with the magnetic field. Since in the absence of warping terms the magnetic field does not open a gap in the dispersion the topological character of the states is preserved (the role of warping is discussed briefly below). $V(x)$ is the electrostatic potential, with the associated electric field $E = - \partial V/\partial x$, and $-e$ the electron charge. The random magnetic impurity potential $U = J s \cdot \sum_{i} S_i \delta(x-x'_i)$ is the contact-like interaction term allowing spin-dependent scattering between an electron and effective magnetic impurities. $s$ indicates the spin operators of the electron and $S_i$ indicates the spin operators of impurities with spin-1/2 sited at position $x_i$. At the end we average over uncorrelated impurities which are all assumed to experience the same exchange interaction $J$ with the mobile carriers. We choose the impurity density and exchange coupling to reproduce the mean free path observed experimentally in the diffusive regime. The local moments are in thermal equilibrium, relaxing their energy and angular momentum rapidly to an external bath \cite{Ref_62}, which corresponds to what is seen experimentally. If the local moments coupled only to the edges the moments on each edge would polarize quickly and the edges would become ballistic in a short amount of time: this is not observed in experiment. The Zeeman interaction between the impurities and the out-of-plane magnetic field $U_Z = g_1 \mu_B  \sum_{i} S_i \cdot B $. The notation $\sigma_z$ in the full Hamiltonian represents the $z$-component of the Pauli spin matrix for a spin-1/2 particle. We focus on the DC limit, where $\omega\tau \ll 1$.

\begin{figure}[tbp]
\includegraphics[width=\columnwidth]{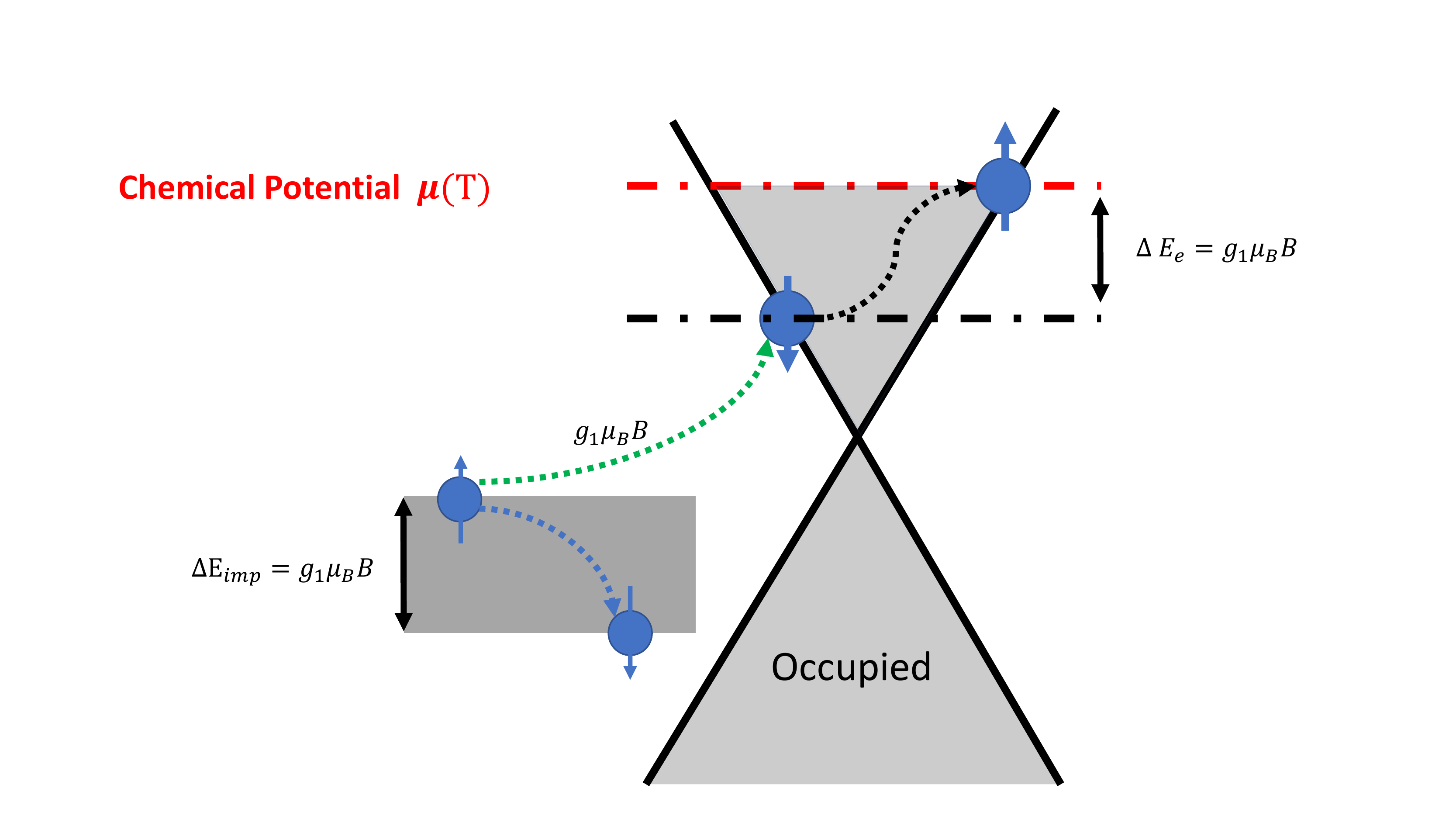}
\caption{Spin-flip scattering. An electron with spin down is scattered into a spin-up state, which, due to spin-momentum locking, travels in the opposite direction; the energy change is given by the impurity Zeeman splitting $E_{1,Z} = g_1 \mu_B B$.}
\label{Figure: Spin-flip scattering}
\end{figure}

\begin{figure}[tbp]
\includegraphics[width=\columnwidth]{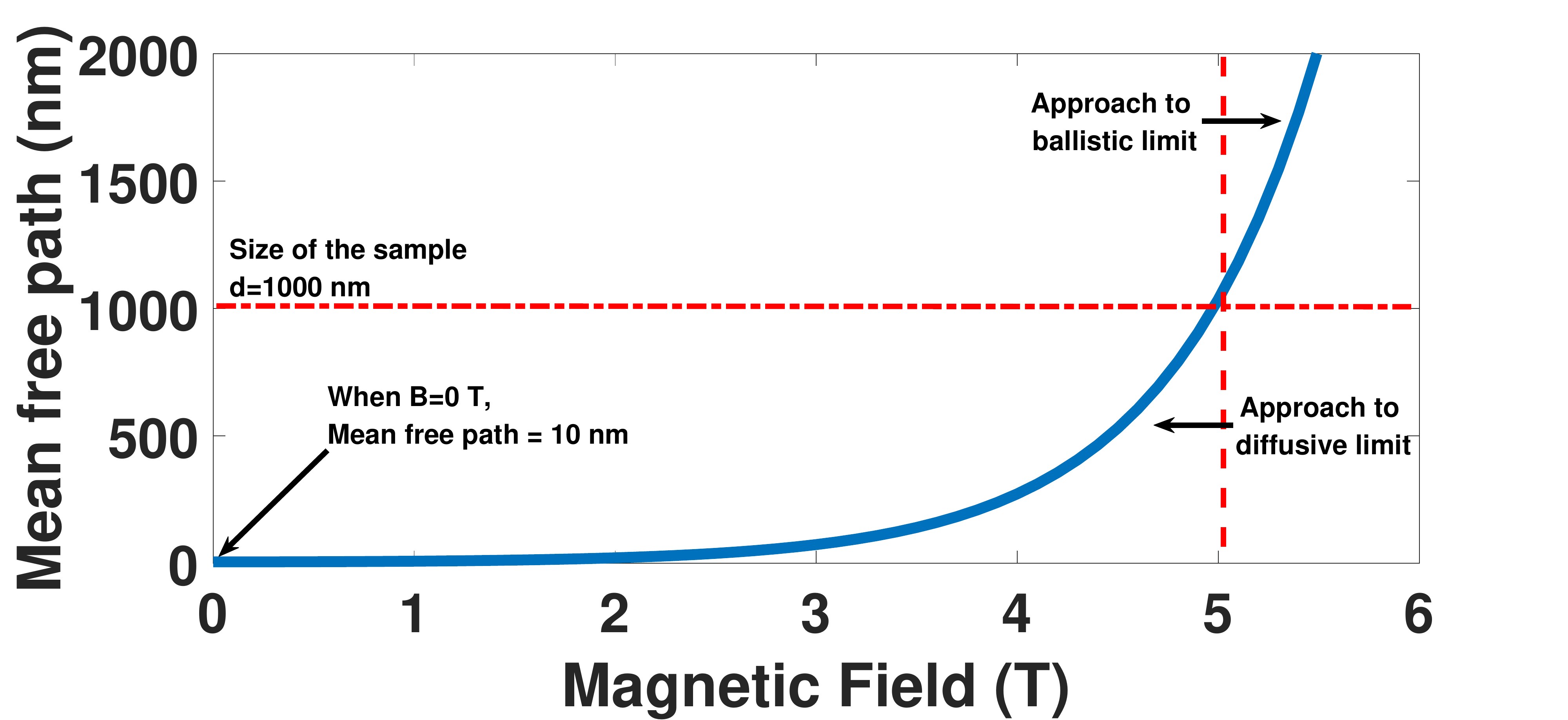}
\caption{The mean free path as a function of magnetic field. The red dashed lines mark the size of the sample $d=1000$nm compared with the mean free path the system. When the magnetic field is small, the system is diffusive, however a larger magnetic field will enhance the mean free path, driving the system into the ballistic regime, leading to a vanishing non-linear response. In all the plots, we have set the mean free path at zero magnetic field to be 10nm.}
\label{Figure: Mean free path} 
\end{figure}

The negative magneto-resistance in Fig.~\ref{Figure: Four-panel conductivity plots} is explained by the relationship between the magnetic field and the mean free path. Although conventional 1D systems are either ballistic or localised, the notion of a mean free path, defined explicitly below, can be applied meaningfully to 1D topological edge states, a diffusive system in which localisation is nevertheless not expected due to topological protection. We note that up and down spins have the same mean free path $l$. Figure \ref{Figure: Spin-flip scattering} gives a diagrammatic example of spin-flip scattering, showing a spin-down electron being flipped to the spin-up channel due to scattering off an impurity. The energy required for this transition is set by the Zeeman splitting of the impurity spin states. As the magnetic field increases the energy cost likewise increases and the transition is suppressed. Figure \ref{Figure: Mean free path} shows the mean free path $l$ increasing as a function of ${\bm B}$ until it exceeds the size $d$ of the sample. Based on this we define the diffusive regime as $l \ll d$, and the ballistic regime as $l > d$. We focus on these two limiting cases, in which simplifying approximations can be made. Specifically, in the diffusive regime one may assume a constant electric field across the channel, and the conductance takes the simple general form $G = \frac{e^2}{h}\, \frac{l}{d}$, where the entire magnetic field dependence is contained in the mean free path $l(B)$. In the ballistic regime it is straightforward to express the current as a function of the potential difference between the source and drain electrodes, and the potential drop occurs overwhelmingly in the vicinity of the electrodes due to contact resistance, although the exact potential profile is immaterial \cite{Ref_1,Ref_2}. The conductance is obtained straightforwardly as $G = \frac{e^2}{h}$. The intermediate region is complicated by potential fluctuations, and is not a focus of current experimental efforts. A full treatment requires accounting for screening thoroughly \cite{Ref_47,Ref_12}.

A perpendicular magnetic field breaks mirror symmetry and enables a second-order response, which increases as a function of the magnetic field in the diffusive regime due to the reduced efficiency of impurity scattering. Nevertheless, the non-linear response vanishes in the ballistic regime. Once transport becomes ballistic there is no more scattering and the impurities become irrelevant. The effective Hamiltonian becomes simply that of a Dirac cone, $H_0 + H_Z$, whence $B$ can be removed by redefining the origin. The second-order response therefore probes the edge state dispersion: if a non-linear response is detected in the ballistic regime it must come from band structure terms of higher order in the wave vector, which are challenging to calculate computationally for 1D systems. Although they can be determined by symmetry their magnitude is generally unknown \cite{Ref_56} (the details are reserved for a future publication).

The response of the other edge can be found by reflecting the Hamiltonian in the $xz$-plane. The Hamiltonian describing the dispersion for the other edge reads $H_0 = -\hbar v_F k_x \sigma_z$. When the magnetic field is flipped the conductance $G$ does not change, consistent with Onsager symmetry. But the direction of the non-linear response on each edge is set by the spin orientation with respect to the magnetic field and the solution to the second order quantum kinetic equation changes sign for the other edge. Hence $\chi$ changes sign, ensuring time reversal breaking in the non-linear response function \cite{Ref_6}.

We derive a quantum kinetic equation following the procedure of Refs.~\onlinecite{Ref_73, Ref_43}, which ensures the Pauli blocking terms are correctly accounted for \cite{Ref_73}. The full details are provided in the Supplemental material (SM) \footnote{The Supplemental material discusses, i) the full form of the quantum kinetic equation, ii) proof for the vanishing elastic scattering, iii) matrix element corresponds to the interaction with impurities, iv) calculation for the other edge, and v) scattering term with the addition of the tilt in the Hamiltonian.}. The system is described by the density matrix $\rho$, which satisfies the quantum Liouville equation $\partial \rho/ \partial t + (i/ \hbar)  [H, \rho] = 0$. The explicit position dependence must be taken into account due to the finite size of the sample. Following a Wigner transformation \cite{Ref_26,Ref_17}
\begin{equation}\label{EQ: quantum kinetic equation}
  \pd{\rho}{t} + \frac{1}{2\hbar} \left\{ \pd{H_0}{k_x}, \pd{\rho}{x} \right\} + J(\rho) = - \frac{e}{\hbar}\pd{V}{x} \pd{\rho}{k_x} 
\end{equation}
In Eq.~\eqref{EQ: quantum kinetic equation}, the single particle density matrix $\rho$ takes the form diag$\{f_\uparrow, f_\downarrow\}$. We write $f_\uparrow = f^{(0)}_\uparrow + g_\uparrow$ is the non-equilibrium distribution for the spin-up electrons composed of the equilibrium part $f^{(0)}_\uparrow$ and out-of-equilibrium part $g_\uparrow$; similarly, $f_\downarrow = f^{(0)}_\downarrow + g_\downarrow$ is the non-equilibrium distribution for the spin-down electrons, the equilibrium distribution have the form $f^{(0)}(\varepsilon) = \left[1+\exp(\beta(\varepsilon - \mu))\right]^{-1}$ where $\beta = (k_B T)^{-1}$. The last term in Eq.~\eqref{EQ: quantum kinetic equation} is the scattering term in the Born approximation, which take the form
\begin{small}
\begin{align}\label{EQ: Scattering terms}
    J(g_\uparrow) =&  \int \big[ P_{k\downarrow,k'\uparrow} f'_\downarrow(1-f_\uparrow) - P_{k\uparrow,k'\downarrow} f_\uparrow(1-f'_\downarrow) \big] \frac{dk'}{2\pi} \\
    J(g_\downarrow) =& \int \big[ P_{k\uparrow,k'\downarrow} f'_\uparrow(1-f_\downarrow) - P_{k\downarrow,k'\uparrow} f_\downarrow(1-f'_\uparrow) \big] \frac{dk'}{2\pi} 
\end{align}
\end{small}
Here $P^i(k', \downarrow \to k, \uparrow)$ indicates the probability of spin-flip scattering between a spin-up electron at $k$ and an impurity, ending with a spin-down electron at $k'$. Primed quantities indicate the final state following a scattering event. We obtain two coupled Boltzmann equations for the spin-up and spin-down electrons:
\begin{small}
\begin{align}\label{EQ: Coupled differential equations 1}
	& \pd{g_\uparrow(\varepsilon)}{x} + \Gamma_1(\varepsilon) g_\uparrow(\varepsilon) - \Gamma_2(\varepsilon) g_\downarrow(\varepsilon_-) = -e \pd{V}{x} \pd{f_\uparrow^{(0)}(\varepsilon)}{\varepsilon} \\\label{EQ: Coupled differential equations 2}
	& \pd{g_\downarrow(\varepsilon)}{x} - \Gamma_1(\varepsilon) g_\uparrow(\varepsilon_+) + \Gamma_2(\varepsilon_+) g_\downarrow(\varepsilon) = -e \pd{V}{x} \pd{f_\downarrow^{(0)}(\varepsilon)}{\varepsilon}
\end{align}
\end{small}
where $\varepsilon_- = \varepsilon - \varepsilon_Z$ and $\varepsilon_+ = \varepsilon + \varepsilon_Z$  The two scattering rates are defined as follows:
\begin{small}
\begin{align}\label{EQ: Scattering rate 1}
	\Gamma_1(\varepsilon) =& \frac{N_i J^2}{\hbar^2 v_F}\bigg[ \frac{1}{1+e^{-\alpha}} \bigg[1-f_\downarrow^{(0)}(\varepsilon)\bigg] +  \frac{1}{1+e^{\alpha}} f_\downarrow^{(0)}(\varepsilon) \bigg] \\\label{EQ: Scattering rate 2}
	\Gamma_2(\varepsilon) =& \frac{N_i J^2}{\hbar^2 v_F} \bigg[ \frac{1}{1+e^{-\alpha}} f_\uparrow^{(0)}(\varepsilon) + \frac{1}{1+e^{\alpha}} \bigg[1-f_\uparrow^{(0)}(\varepsilon)\bigg] \bigg],
\end{align}
\end{small}
where $N_i$ is the number of impurities, the dimensionless factor $\alpha = g_1 \mu_B B / (k_BT)^{-1}$, and the change of the Zeeman energy during the spin-flipping interactions $\varepsilon_Z = g_1 \mu_B B$. We solve the coupled Eq.~\eqref{EQ: Coupled differential equations 1} and Eq.~\eqref{EQ: Coupled differential equations 2} by integrating separately over left and right movers, which also ensures the correct solution in the ballistic regime:
\begin{widetext}
\begin{align}\label{EQ: First order solution 1}
	g^{(1)}_\uparrow =& -e \int_0^x \bigg[\frac{\Gamma_2}{\kappa} + \frac{\Gamma_1}{\kappa} \exp[\kappa(x-x')]\bigg] \pd{V}{x} \pd{f_\uparrow^{(0)}}{\varepsilon} dx' + e\int_d^x  \bigg[\frac{\Gamma_2}{\kappa}\big(\exp[\kappa(x-x')]-1\big) \bigg] \pd{V}{x} \pd{f_\downarrow^{(0)}}{\varepsilon} dx' \\\label{EQ: First order solution 2}
	g^{(1)}_\downarrow =& -e \int_d^x \bigg[\frac{\Gamma_1}{\kappa} + \frac{\Gamma_2}{\kappa} \exp[\kappa(x-x')] \bigg] \pd{V}{x} \pd{f_\downarrow^{(0)}}{\varepsilon} dx' + e \int_0^x \bigg[ \frac{\Gamma_1}{\kappa}  \big( \exp[\kappa(x-x')]-1\big) \bigg] \pd{V}{x} \pd{f_\uparrow^{(0)}}{\varepsilon} dx'
\end{align}
\end{widetext}
where $\kappa = \Gamma_1+\Gamma_2$, and the mean free path $l = \kappa^{-1} = (\Gamma_1+\Gamma_2)^{-1}$. The current density $j = -e {\rm Tr}(v\rho)$, where $v = (1/\hbar)(\partial H_0/\partial k)$. For spin-up electrons the momentum integration is performed over $k>0$, and for spin-down electrons over $k<0$. For the second-order response the sign of the magnetic field in the scattering terms Eq.~\eqref{EQ: Scattering rate 1}, \eqref{EQ: Scattering rate 2} will change, yet $\Gamma_1$ and $\Gamma_2$ are symmetric in $\alpha$; the sign of the driving term will change similarly to the band dispersion. The mean free path is unchanged and the formal solution to the differential equation is analogous to Eqs.~\ref{EQ: First order solution 1}-\ref{EQ: First order solution 2}, with the replacements $\pd{f^{(0)}}{\varepsilon} \rightarrow \pd{g^{(1)}}{\varepsilon}$. These equations cannot be reduced to a simple closed form and are solved iteratively.

In this work we have not discussed dispersions beyond the linear case. In Bi$_2$Se$_3$ warping complicates the dispersion, and in the spin eigenstate basis $H_0 = A k_x \sigma_y + C \sigma_z k_x^3$. A magnetic field $\parallel \hat{\bm y}$ yields a term of the form $\sigma_z B_y$ due to warping, which opens a small gap in the edge spectrum. However, warping only accounts for up to 10$\%$ of the Fermi energy, thus the gap is expected to be small, and will not influence the dynamics in the vicinity of the Fermi energy discussed here. Since the addition of warping complicates significantly the description of the interaction with the impurities, a further derivation is beyond the scope of this paper.

In summary, we have shown that a magnetic field drastically enhances the conductivity of topological edge states and gives rise to an edge-dependent non-linear response which vanishes in the ballistic limit. The magnetic field, as well as proximity to a ferromagnet, can be used to drive the system into the ballistic regime, while the non-linear response probes the edge state dispersion. In the future the transport theory can be extended to the Kondo regime along the lines of \onlinecite{Ref_43}. 

\textit{Acknowledgments}. This research is supported by the Australian Research Council Centre of Excellence in Future Low-Energy Electronics Technologies (project CE170100039) and funded by the Australian Government. PB acknowledges the National Key Research and Development Program of China (grant No. 2017YFA0303400), China postdoctoral science foundation (grant no. 2019M650461) and NSFC grant no. U1930402 for financial support.
%

\end{document}